\documentclass[10pt,conference,final]{IEEEtran}
\usepackage[top=1.5cm, bottom=1.5cm, left=1.5cm, right=1.5cm]{geometry}

\usepackage[british]{babel}
\usepackage[hyphens]{url}
\usepackage{paralist}
\usepackage{graphicx}
\usepackage[noadjust]{cite}
\usepackage{siunitx}
\usepackage[pdftex,colorlinks=true]{hyperref}

\IEEEoverridecommandlockouts

\begin{document}

\title{``Can I Implement Your Algorithm?'':\\A Model for Reproducible Research Software}

\author{\IEEEauthorblockN{Tom Crick\thanks{Tom Crick would like to
      acknowledge financial support from the Software Sustainability
      Institute as a 2014 Fellow.}}
\IEEEauthorblockA{Department of Computing\\
Cardiff Metropolitan University\\
Cardiff, UK\\
Email: {\url{tcrick@cardiffmet.ac.uk}}}
\and
\IEEEauthorblockN{Benjamin A. Hall and Samin Ishtiaq}
\IEEEauthorblockA{Microsoft Research\\
Cambridge, UK\\
Email: {\url{{benhall,samin.ishtiaq}@microsoft.com}}}}

\maketitle

\begin{abstract}
The reproduction and replication of novel results has become a major
issue for a number of scientific disciplines. In computer science and
related computational disciplines such as systems biology, the issues
closely revolve around the ability to implement novel algorithms and
approaches. Taking an approach from the literature and applying it to
a new codebase frequently requires local knowledge missing from the
published manuscripts and project websites. Alongside this issue,
benchmarking, and the development of fair --- and widely available ---
benchmark sets present another barrier.

In this paper, we outline several suggestions to address these issues,
driven by specific examples from a range of scientific domains.
Finally, based on these suggestions, we propose a new open platform
for scientific software development which effectively isolates
specific dependencies from the individual researcher and their
workstation and allows faster, more powerful sharing of the results of
scientific software engineering.
\end{abstract}

\IEEEpeerreviewmaketitle

\section{Introduction}

Marc Andreessen (co-author of Mosaic, the first widely used Web
browser) famously said in 2011 that ``{\emph{software is eating the
world}}''~\cite{andreessen:2011}. It is true: we clearly live in a
computational world, with our everyday communications, entertainment,
shopping, security, banking, transportation, etc, all heavily
dependent on (or replaced by) software.

This is particularly true for science and engineering. A 2012 report
by the Royal Society stated that computational techniques have
``{\emph{moved on from assisting scientists in doing science, to
transforming both how science is done and what science is
done}}''~\cite{rssaaoe:2012}. New experiments, simulations, models,
benchmarks, even proofs cannot be done without software. And this
software does not consist of simple hack-together, use-once,
throw-away scripts; scientific software repositories contain
thousands, perhaps millions, of lines of code and they increasingly
need to be actively supported and maintained. More importantly, with
reproducibility being a fundamental tenet of science, they need to be
re-useable.

However, if we closely analyse the scientific literature related to
software tools it often does not appear to be adhering to these
rules~\cite{nature:2011}. How many of them are reproducible? How many
explain their experimental methodologies, in particular the basis for
their benchmarking? In particular, can we (re)build the
code~\cite{collberg-et-al:2014}? We, the authors, are perhaps as
guilty as anyone in the past, where we have published
papers~\cite{crick-et-al:2009a,Berdine2011SLAyer} with benchmarks and
promises of code to be released in the near future.

There are numerous reasons why the wider scientific community is in
this state. We are experiencing significant changes in academic
dissemination and publication, especially the open access movement,
with new models being
proposed~\cite{stodden-et-al:2013,fursin+dubach:2014}. It is partly
cultural: there are numerous non-technical impediments to making
software maintainable and re-useable, too. The pressure to ``make the
discovery'' and publish quickly disincentivises careful software
curation. Releasing code prematurely is often seen to give your
competitors an advantage, but we should be shining light into these
``black boxes''~\cite{morin-et-al:2012}. In essence: better software,
better research~\cite{goble:2014}.

Nevertheless, there has been previous work in this
area~\cite{sim-et-al:2003,chirigati-et-al:2013}, as well as a range of
manifestos for reproducible research and community initiatives, such
as the Recomputation
Manifesto~\cite{gent:2013}\footnote{\url{http://www.recomputation.org/}}
and
cTuning~\cite{fursin-et-al:2014}\footnote{\url{http://ctuning.org/}},
along with curated recommendations on where to publish research
software\footnote{\url{http://www.software.ac.uk/resources/guides/which-journals-should-i-publish-my-software}}.

However, things can, should and need to be much better. In this paper,
we present a call to action, along with a set of recommendations which
we hope will lead to better, more sustainable, more re-useable
software, to move towards an imagined future practice of software
development and usage in science and engineering.  The basis for many
of these recommendations is predicated on the basic scientific tenet
of openness.

\section{A Model for Reproducible Research Software}

\subsection{Can I Implement Your Algorithm?}

Reproducibility is a fundamental tenet of good science. Yet many
descriptions of algorithms are too high-level, too obscure, too
poorly-defined to allow an easy re-implementation by a third party. A
step in the algorithm might say: ``{\emph{We pick an element from the
frontier set}}'' but which element do you pick? Will the first one do?
Why will any element suffice? Sometimes the author would like to give
more implementation detail but is constrained by the paper page
limit. Sometimes the authors' description in-lines other algorithms or
data structures that perhaps only that author is familiar with.

\noindent {\textbf{Recommendation {\textrm{I}}:}} We recommend here
that a paper must describe the algorithm in such a way that it is
implementable by any reader of that algorithm. This is subjective, of
course. Therefore, we also recommend that relevant scientific
conferences have a special track for papers that re-implement past
papers' algorithms, techniques or tools, as well as incentives to
support sharing of computational artefacts (for example, the Artifact
Evaluation process as part of the 2014 ACM SIGPLAN Conference on
Object-Oriented Programming, Systems, Languages \&
Applications (OOPSLA)\footnote{\url{http://2014.splashcon.org/track/splash2014-artifacts}}).

\subsection{Set The Code Free} 

There can be no better proof that your algorithm works, than if you
provide the source code of an implementation. Software development is
hard, but sharing and re-using code is relatively easy.

Many years ago, Richard Stallman (founder of the GNU Project and Free
Software Foundation) postulated that all code would be
free~\cite{rms:2010} and we would make our money by consulting on the
code.  As it turns out, this is now the case for a significant
part of the computing industry. There are, of course, hard commercial
pressures for keeping code closed-source. Even in the scientific
domain, scientists and their collaborators may wish to hold onto their
code as a competitive advantage, especially if there exists larger
competitors who could use the available code to ``reverse scoop'' the
inventors, charging into a promising new research area opened by the
inventors.

Closed source is one thing. Licenses that deny the user from viewing,
modifying, or sharing the source are another thing. There are,
however, even licences on widely adopted tools like
GAUSSIAN~\cite{Giles2004} that prohibit even analysing software
performance and behaviour. For example, a wide variety of licenses
exist for molecular dynamics software, with different degrees of
openness (GROMACS uses the GNU Lesser General Public License
(LGPL)~\cite{Hess2008}, CHARMM and Desmond are Academic/Commercial
software licences~\cite{Brooks2009,Bowers2006}, Amber and NAMD are
custom open-like licences). Z3 is an example from the verification
area: the code itself is not open source, but the MSR-LA license that allows
the source code to be read, copied, forked for academic use, provides
researchers in the field much more than
before~\cite{deMoura2012Z3open}.

\noindent {\textbf{Recommendation {\textrm{II}}:}} There is little
doubt that, if science wants to be open and free, then the code that
underlies it too needs to be open and free. Code that is available for
browsing, modifying, and forking facilitates testing and comparison,
and promotes competition. We recommend that code be published under an
appropriate open source license~\cite{osl}; while we defer legal
discussion of the specifics of any particular licences, BSD and Apache are good,
flexible ones.

Ultimately: set the code free. Put it on a public space such as
GitHub, where it is easy to share and fork. You should embrace the
spirit of the (somewhat tongue-in-cheek) CRAPL academic-strength open
source license\footnote{\url{http://matt.might.net/articles/crapl/}}
and publish your code -- it is good enough~\cite{barnes:2010}.

\subsection{Be A Better Person}

If you have the appropriate skills and the experience, you can always
create better software. We have seen the emergence of successful
initiatives, such as the Software Sustainability
Institute\footnote{\url{http://www.software.ac.uk/}}, Software
Carpentry\footnote{\url{http://software-carpentry.org/}} and the UK
Community of Research Software
Engineers\footnote{\url{http://www.rse.ac.uk}}, in cultivating
world-class research through software, developing software skills and
raising the profile of research software engineers.

Many scientists will not have had any formal, or even informal,
training in scientific software development. Even basic training in software engineering
concepts like version control, unit testing, build tools, etc, can help improve the
quality of the software written enormously~\cite{wilson2006}.
Interestingly, many of these concepts are taught to computer science
undergraduates, but it could be argued that they are taught at the
wrong time of their careers, without the experience of complex,
long-running projects.

\noindent {\textbf{Recommendation {\textrm{III}}:}} Software
development skills should be regarded as fundamental literacies for
scientists and engineers: we recommend that basic programming and
computational skills are taught as core at undergraduate and
postgraduate level.

\subsection{Latin Is The Language Of God} 

There is no other scientific or technical field where its participants
can just make up a non-principled artefact like a programming language
so easily. In a way, it shows how much of a ``commons'' computer
science has become, that anyone can create a new programming language,
API, framework or compiler. This clearly has its advantages and
disadvantages.

High-level languages are generally more readable than their
competitors. The ``density'' of a program is often seen to be a good
thing, but it is not always the case that a shorter Haskell program
(for example) is easier to maintain than a longer Python/C++
one. Nevertheless, what is important is the readability of the code
itself. A good example here is from the world of automatic theorem
proving: the SSReflect language is much more readable than the
original, standard Coq language~\cite{GonthierZND13}. SSReflect uses
mathematicians' vernacular for script commands, allows reproducibility
of automatic proof-checking because parameters are named rather than
numbered.  Even though these proof scripts are really only ever going
to be run by a machine, they seek to maintain the basic mathematical
idea that a proof should be readable by another mathematician.

High-level programming languages impose constraints like types: that you can
never add a number and a string is the most basic example, but ML's
functors provide principled ways of plugging in components with their
implementations completely hidden. Aggressive type checking avoids a
subset of bugs which can arise due to incorrectly written functions
e.g. well publicised problems with a NASA Mars orbiter\footnote{See:
\url{http://www.cnn.com/TECH/space/9909/30/mars.metric.02/}}.  A
further example is a pressure coupling
bug\footnote{\url{http://redmine.gromacs.org/issues/14}} in
GROMACS~\cite{Hess2008}, which arose due to the inappropriate swapping
of a pressure term with a stress tensor.  A further extension of
types, a concept called units of measure that is implemented in
languages such as F\#, can deal with these kinds of bugs at compile
time. Similarly, problems found using in-house software for
crystallography led to the retraction of five
papers~\cite{Miller2006}, due to a bug which inverted the phases.

\noindent {\textbf{Recommendation {\textrm{IV}}:}} The use of a
principled, high-level programming language in which to write your
software helps hugely with the maintainability, robustness and
openness of the software produced.

\subsection{Test It To See}

Some models may be chaotic and influenced by floating-point errors
(e.g. molecular dynamics), further frustrating testing. For example:
Sidekick is an automated tool for building molecular models and
performing simulations~\cite{Hall2014Sidekick}. Each system is
simulated from an different initial random seed, and under most
circumstances this is the only difference expected between
replicas. However, on a mixed cluster with both AMD and Intel
microprocessors on the nodes, the difference in architecture was found
to alter the number of water molecules added to each system by
one. This meant that the same simulation performed on different
architectures would diverge. Similarly, in a different simulation
engine, different neighbour searching strategies gave divergent
simulations due to the differing order in which forces were summed.

\noindent {\textbf{Recommendation {\textrm{V}}:}} Despite these
challenges to testing, unshared code is ultimately untestable.
Testing new complex scientific software is difficult -- until the
software is complete, unit tests may not be available. You should thus
aim to link to/from publicly-shared code: shared code is inherently more
test-able.

\subsection{Lineage (or: ``Standing On The Shoulders Of Giants'')} 

Research software is not just software -- it is the instantiation of
novel algorithms and data structures (or at least novel applications
of data structures). Thus, lineage is important:

\noindent {\textbf{Recommendation {\textrm{VI}}:}} Code should always
include links to papers publishing key algorithms and the code should
include explicit relationships to other projects on the repository
(i.e. {\emph{Project B}} was branched from {\emph{Project A}}). This
ensure that both the researchers and software developers working
upstream of the current project are properly credited, encouraging
future sharing and development. Remember, the people who did the
research are not necessarily the same people as the developers and
maintainers of the software, so it is important to reward both
appropriately with citations (a good way of doing this is the use of
CITATION
files\footnote{\url{http://blog.rtwilson.com/encouraging-citation-of-software-introducing-citation-files/}}).

\subsection{YMMV}

\begin{figure}[!ht]
\centering
\includegraphics[width=0.9\columnwidth]{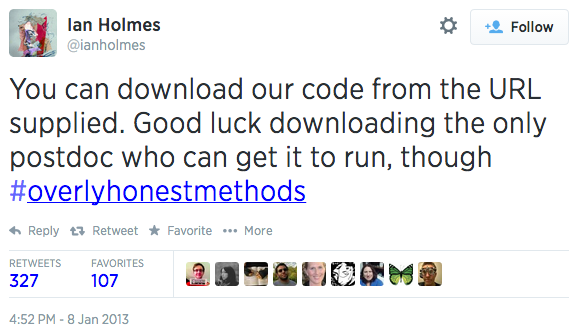}
\caption{{\texttt{\#overlyhonestmethods}} on Twitter\newline [source: \url{https://twitter.com/ianholmes/status/288689712636493824}]}
\label{fig:overlyhonestmethod} 
\end{figure}

The tweet in Figure~\ref{fig:overlyhonestmethod} is sad but worryingly
true, highlighting the perils of reproducible research\footnote{Also
see: \url{http://www.phdcomics.com/comics.php?f=1689}}. Often, the
tool that the paper describes does not exist for download. Or runs
only on one particular bespoke platform. Or might run for the author,
for a while, but will `bit-rot' so quickly that even the author cannot
compile it in a couple of month's time.

\noindent {\textbf{Recommendation {\textrm{VII}}:}} Providing the
source code of the tool helps, of course. But you must also provide
details of precisely \emph{how} you built and wrote the software. For
example:

\begin{compactitem}
\item you should provide the compiler and build toolchain; 
\item you should provide build tools (e.g. Makefiles/Ant/etc) and
  comprehensive build instructions; 
\item you should list or link to all non-standard packages and libraries that you use; 
\item you should note the specifics of the hardware and OS used. 
\end{compactitem}

This may appear to be significant extra overhead for researchers, but
GitHub APIs, continuous integration servers, virtual machines and
cloud environments can make it easier; see
Section~\ref{sec:Conclusion} for more on this.

\subsection{Data Representations and Formats}

We often do not, and should not, care how things are stored on disk,
what their precise representations are. But a common, constrained,
standard representation is good for passing tests or models around
between different tools. A properly described representation, like the
SMT-LIB format\footnote{\url{http://smt-lib.org}} for Satisfiability
Modulo Theory (SMT) solvers, where both the syntax and semantics are
well understood, hugely aids developing tools, techniques and
benchmarks.

Another example, from biology, is that of the standard representation
of qualitative networks and Boolean
networks~\cite{Kauffman1969,Schaub2007}.  These networks can be
expressed in SMV format, but this would mean that standard
qualitative/Boolean network behaviours have to be hard-coded for each
variable, introducing the possibility for errors. In the
BioModelAnalyzer tool~\cite{Benque2012}, the XML contains \emph{only}
the modifiable parameters limiting the possibility for error.

\noindent {\textbf{Recommendation {\textrm{VIII}}:}} Avoid creating
new representations when common formats already exist. Use existing
extensible internationally standardised representations and formats to
facilitate sharing and re-use.

\subsection{World Records}

The benchmarks the tool describes are fashioned only for this instance
of this time. They might claim to be from the Windows device driver
set, but the reality is that they are stripped down versions of the
originals. Stripped down so much as to be useless to anyone but the
author vs. the referee. It is worse than that really: enough
benchmarks are included to beat other tools. The comparisons are never
fair (neither are other peoples' comparisons against your tool). If
every paper has to be novel, then every benchmark, too, will be novel;
there is no monotonic, historical truth in new, synthetically-crafted
benchmarks. It is as if, in order to beat Usain Bolt's
\num{100}\si{\metre} world record time, you make him wear boots on a
muddy icy track, weighing him down with \num{50}\si{\kilogram} of
excess weight. Given this set up, you could surely hope to beat his
\num{9.63}\si{\second} time on a shorter length track.

\noindent {\textbf{Recommendation {\textrm{IX}}:}} Benchmarks should
be public. They should allow anyone to contribute, implying that the
tests are in a standard format. Further, these benchmarks must be
heavily curated. Every test/assertion should be justified. Papers
should be penalised if they do not use these public benchmarks. While
there are some domains in which it may not be immediately possible to
share full benchmarks sets, this should be the exception (with
justification) rather than the norm.

A good example of some of these points is the RCSB Protein Data
Bank\footnote{\url{http://www.pdb.org}} and Systems Biology Markup
Language~\cite{Chaouiya2013}. The software ones we know of,
the SMT
Competition\footnote{\url{http://smtcomp.sourceforge.net/2014/}},
SV-COMP\footnote{\url{http://sv-comp.sosy-lab.org/2015/}} and
Termination Problems Data
Base\footnote{\url{http://termination-portal.org/wiki/TPDB}} are on
that journey. Such repositories would allow the tests to be taken and
easily analysed by any competitor tool.

\subsection{Welcome to Web 2.0}

Virtual machines (VMs) in the cloud also make the testing of scaling
properties more simple.  If you have a tool that you claim is more
efficient, you could put together a cluster of slow nodes in the cloud
to demonstrate how well the software scales for parallel calculations.
Cloud computing is cheap, and getting cheaper. Algorithms that used to
require massive HPC resources can now be run cheaply by bidding on the
VM spot market. The Web is a great leveller: use and share workflows
and web services~\cite{crick-et-al:2009b}.

\noindent {\textbf{Recommendation {\textrm{X}}:}} The Web and the
cloud really do open up a whole new way of working. Even small,
seemingly trivial features like putting up a web interface to your
tool and its tests will allow users who are not able to install
necessary dependencies to explore the running of the tool
\cite{Hall2014}. Ultimately, this can lead to making an ``executable
paper'' appear on the Internet. The interactive {\em Try
F\#}\footnote{\url{http://www.tryfsharp.org/Learn}} and Z3
tutorials\footnote{\url{http://rise4fun.com/Z3/tutorial/guide}} are a
great start that begin to expose what can be done in this area.

\section{Conclusions: A New Model}\label{sec:Conclusion} 

\noindent{\textbf{This is how we imagine the future for research
software:}}\newline\newline Suppose you have come up with a better
algorithm to deal with some standard problem.  You write up the paper
on the algorithm, and you also push a C++ implementation of your
algorithm to the our cloud environment's section on this standard problem.

The effect of pushing your implementation is to register your program
as a possible competitor in this standard problem competition. There
are several dozen widely-agreed tests on this problem already on our
cloud environment's database. Maybe, after some negotiation due to
your novel approach to this standard problem, you add some of your own
tests to the database too.

Pushing your code activates the environment's continuous integration
system.  The cloud pulls in all the dependencies your code needs, on
the platforms you specify, and runs all the benchmarks. This happens
every time you push. It also happens every time one of your
dependencies (a library, a firmware upgrade for your platform, a new
API) changes too.

If we are truly serious about addressing the systemic socio-technical
issues in scientific disciplines that are underpinned by leveraging
software and computational techniques, then the proposal above would
bring together almost all of the points we have discussed in this
paper to provide an open research infrastructure for all. There are
already several web services that nearly do a number of part of
this. Something more complete, and stamped with the authority of the
major domain conferences/journals/professional societies, would mean
that your code would never `bit-rot', and no one would have problems
reproducing the implementation of your published algorithm.

\IEEEtriggeratref{27}

\bibliographystyle{IEEEtran}
\bibliography{wssspe2}

\begin{thebibliography}{10}
\providecommand{\url}[1]{#1}
\csname url@samestyle\endcsname
\providecommand{\newblock}{\relax}
\providecommand{\bibinfo}[2]{#2}
\providecommand{\BIBentrySTDinterwordspacing}{\spaceskip=0pt\relax}
\providecommand{\BIBentryALTinterwordstretchfactor}{4}
\providecommand{\BIBentryALTinterwordspacing}{\spaceskip=\fontdimen2\font plus
\BIBentryALTinterwordstretchfactor\fontdimen3\font minus
  \fontdimen4\font\relax}
\providecommand{\BIBforeignlanguage}[2]{{%
\expandafter\ifx\csname l@#1\endcsname\relax
\typeout{** WARNING: IEEEtran.bst: No hyphenation pattern has been}%
\typeout{** loaded for the language `#1'. Using the pattern for}%
\typeout{** the default language instead.}%
\else
\language=\csname l@#1\endcsname
\fi
#2}}
\providecommand{\BIBdecl}{\relax}
\BIBdecl

\bibitem{andreessen:2011}
M.~Andreessen, ``{Why Software Is Eating The World},'' \emph{{The Wall Street
  Journal}}, August 2011, available online:
  \url{http://online.wsj.com/news/articles/SB10001424053111903480904576512250915629460}.

\bibitem{rssaaoe:2012}
{Royal Society}, ``Science as an open enterprise,'' 2012, available from:
  \url{https://royalsociety.org/policy/projects/science-public-enterprise/report/}.

\bibitem{nature:2011}
Editorial, ``Devil in the details,'' \emph{Nature}, vol. 470, no. 7334, pp.
  305--306, 2011.

\bibitem{collberg-et-al:2014}
C.~Collbery, T.~Proebsting, G.~Moraila, A.~Shankaran, Z.~Shi, and A.~M. Warren,
  ``{Measuring Reproducibility in Computer Systems Research},'' Department of
  Computer Science, University of Arizona, Tech. Rep., 2014.

\bibitem{crick-et-al:2009a}
T.~Crick, M.~{De Vos}, M.~Brain, and J.~Fitch, ``{Generating Optimal Code using
  Answer Set Programming},'' in \emph{{Proceedings of 10th International
  Conference on Logic Programming and Nonmonotonic Reasoning (LPNMR'09)}}, ser.
  Lecture Notes in Computer Science, vol. 5753.\hskip 1em plus 0.5em minus
  0.4em\relax Springer, 2009, pp. 554--559.

\bibitem{Berdine2011SLAyer}
J.~Berdine, B.~Cook, and S.~Ishtiaq, ``{SLAyer: Memory Safety for Systems-Level
  Code},'' in \emph{{Proceedings of the 23rd International Conference on
  Computer Aided Verification (CAV 2011)}}, ser. Lecture Notes in Computer
  Science, vol. 6806.\hskip 1em plus 0.5em minus 0.4em\relax Springer, 2011,
  pp. 178--183.

\bibitem{stodden-et-al:2013}
V.~Stodden, P.~Guo, and Z.~Ma, ``{Toward Reproducible Computational Research:
  An Empirical Analysis of Data and Code Policy Adoption by Journals},''
  \emph{{PLoS ONE}}, vol.~8, no.~6, 2013.

\bibitem{fursin+dubach:2014}
G.~Fursin and C.~Dubach, ``{Community-Driven Reviewing and Validation of
  Publications},'' in \emph{{Proceedings of the 1st ACM SIGPLAN Workshop on
  Reproducible Research Methodologies and New Publication Models in Computer
  Engineering (TRUST'14)}}.\hskip 1em plus 0.5em minus 0.4em\relax ACM Press,
  2014, pp. 1--4.

\bibitem{morin-et-al:2012}
A.~Morin, J.~Urban, P.~D. Adams, I.~Foster, A.~Sali, D.~Baker, and P.~Sliz,
  ``{Shining Light into Black Boxes},'' \emph{Science}, vol. 336, no. 6078, pp.
  159--160, 2012.

\bibitem{goble:2014}
C.~Goble, ``{Better Software, Better Research},'' \emph{{IEEE Internet
  Computing}}, vol.~18, no.~5, pp. 4--8, 2014.

\bibitem{sim-et-al:2003}
S.~Sim, S.~Easterbrook, and R.~Holt, ``Using benchmarking to advance research:
  a challenge to software engineering,'' in \emph{{Proceedings of the 25th
  International Conference on Software Engineering (ICSE 2003)}}.\hskip 1em
  plus 0.5em minus 0.4em\relax IEEE Press, 2003, pp. 74--83.

\bibitem{chirigati-et-al:2013}
F.~Chirigati, M.~Troyer, D.~Shasha, and J.~Freire, ``{A Computational
  Reproducibility Benchmark},'' \emph{{IEEE Data Engineering Bulletin}},
  vol.~36, no.~4, pp. 54--59, 2013.

\bibitem{gent:2013}
I.~P. Gent, ``{The Recomputation Manifesto},'' April 2013, available from:
  \url{http://arxiv.org/abs/1304.3674}.

\bibitem{fursin-et-al:2014}
G.~Fursin, R.~Miceli, A.~Lokhmotov, M.~Gerndt, M.~Baboulin, A.~D. Malony,
  Z.~Chamski, D.~Novillo, and D.~{Del Vento}, ``{Collective mind: Towards
  practical and collaborative auto-tuning},'' \emph{{Scientific Programming}},
  vol.~22, no.~4, pp. 309--329, 2014.

\bibitem{rms:2010}
R.~M. Stallman, \emph{{Free Software Free Society: Selected Essays of Richard
  M. Stallman}}.\hskip 1em plus 0.5em minus 0.4em\relax Free Software
  Foundation, 2010.

\bibitem{Giles2004}
J.~Giles, ``Software company bans competitive users,'' \emph{Nature}, vol. 429,
  no. 6989, 2004.

\bibitem{Hess2008}
B.~Hess, C.~Kutzner, D.~van~der Spoel, and E.~Lindahl, ``{GROMACS 4: Algorithms
  for Highly Efficient, Load-Balanced, and Scalable Molecular Simulation},''
  \emph{Journal of Chemical Theory and Computation}, vol.~4, no.~3, pp.
  435--447, 2008.

\bibitem{Brooks2009}
B.~R. Brooks, C.~L. Brooks, A.~D. Mackerell, L.~Nilsson, R.~J. Petrella,
  B.~Roux, Y.~Won, G.~Archontis, C.~Bartels, S.~Boresch, A.~Caflisch, L.~Caves,
  Q.~Cui, A.~R. Dinner, M.~Feig, S.~Fischer, J.~Gao, M.~Hodoscek, W.~Im,
  K.~Kuczera, T.~Lazaridis, J.~Ma, V.~Ovchinnikov, E.~Paci, R.~W. Pastor, C.~B.
  Post, J.~Z. Pu, M.~Schaefer, B.~Tidor, R.~M. Venable, H.~L. Woodcock, X.~Wu,
  W.~Yang, D.~M. York, and M.~Karplus, ``{CHARMM: The biomolecular simulation
  program},'' \emph{Journal of Computational Chemistry}, vol.~30, no.~10, pp.
  1545--1614, 2009.

\bibitem{Bowers2006}
K.~J. Bowers, E.~Chow, H.~Xu, R.~O. Dror, M.~P. Eastwood, B.~A. Gregersen,
  J.~L. Klepeis, I.~Kolossvary, M.~A. Moraes, F.~D. Sacerdoti, J.~K. Salmon,
  Y.~Shan, and D.~E. Shaw, ``Scalable algorithms for molecular dynamics
  simulations on commodity clusters,'' in \emph{Proceedings of the 2006
  ACM/IEEE Conference on Supercomputing}.\hskip 1em plus 0.5em minus
  0.4em\relax IEEE Press, 2006.

\bibitem{deMoura2012Z3open}
L.~de~Moura, ``{Releasing the {Z3} source code},'' 2012, available online:
  \url{http://leodemoura.github.io/blog/2012/10/02/open-z3.html}.

\bibitem{osl}
``{Open Source Licenses},'' \url{http://opensource.org/licenses}.

\bibitem{barnes:2010}
N.~Barnes, ``Publish your computer code: it is good enough,'' \emph{Nature},
  vol. 467, no. 753, 2010.

\bibitem{wilson2006}
G.~Wilson, ``Software carpentry: Getting scientists to write better code by
  making them more productive,'' \emph{{Computing in Science \& Engineering}},
  vol.~8, no.~6, 2006.

\bibitem{GonthierZND13}
G.~Gonthier, B.~Ziliani, A.~Nanevski, and D.~Dreyer, ``How to make ad hoc proof
  automation less ad hoc,'' \emph{{Journal of Functional Programming}},
  vol.~23, no.~4, pp. 357--401, 2013.

\bibitem{Miller2006}
G.~Miller, ``{A Scientist's Nightmare: Software Problem Leads to Five
  Retractions},'' \emph{Science}, vol. 314, no. 5807, pp. 1856--1857, 2006.

\bibitem{Hall2014Sidekick}
B.~A. Hall, K.~B.~A. Halim, A.~Buyan, B.~Emmanouil, and M.~S.~P. Sansom,
  ``Sidekick for membrane simulations: Automated ensemble molecular dynamics
  simulations of transmembrane helices,'' \emph{Journal of Chemical Theory and
  Computation}, vol.~10, no.~5, pp. 2165--2175, 2014.

\bibitem{Kauffman1969}
S.~A. Kauffman, ``Metabolic stability and epigenesis in randomly constructed
  genetic nets,'' \emph{Journal of Theoretical Biology}, vol.~22, no.~3, pp.
  437--67, 1969.

\bibitem{Schaub2007}
M.~A. Schaub, T.~A. Henzinger, and J.~Fisher, ``Qualitative networks: a
  symbolic approach to analyze biological signaling networks,'' \emph{{BMC
  Systems Biology}}, vol.~1, p.~4, 2007.

\bibitem{Benque2012}
D.~Benque, S.~Bourton, C.~Cockerton, B.~Cook, J.~Fisher, S.~Ishtiaq,
  N.~Piterman, A.~Taylor, and M.~Y. Vardi, ``{BMA: visual tool for modeling and
  analyzing biological networks},'' in \emph{{Proceedings of the 24th
  International Conference on Computer Aided Verification (CAV 2012)}}, ser.
  Lecture Notes in Computer Science, vol. 7358.\hskip 1em plus 0.5em minus
  0.4em\relax Springer, 2012, pp. 686--692.

\bibitem{Chaouiya2013}
C.~Chaouiya, D.~Berenguier, S.~M. Keating, A.~Naldi, M.~P. van Iersel,
  N.~Rodriguez, A.~Drager, F.~Buchel, T.~Cokelaer, B.~Kowal, B.~Wicks,
  E.~Goncalves, J.~Dorier, M.~Page, P.~T. Monteiro, A.~von Kamp, I.~Xenarios,
  H.~de~Jong, M.~Hucka, S.~Klamt, D.~Thieffry, N.~Le~Novere, J.~Saez-Rodriguez,
  and T.~Helikar, ``{SBML} qualitative models: a model representation format
  and infrastructure to foster interactions between qualitative modelling
  formalisms and tools,'' \emph{{BMC Systems Biology}}, vol.~7, 2013.

\bibitem{crick-et-al:2009b}
T.~Crick, P.~Dunning, H.~Kim, and J.~Padget, ``{Engineering Design Optimization
  using Services and Workflows},'' \emph{{Philosophical Transactions of the
  Royal Society A: Mathematical, Physical and Engineering Sciences}}, vol. 367,
  no. 1898, pp. 2741--2751, 2009.

\bibitem{Hall2014}
B.~A. Hall, E.~Jackson, A.~Hajnal, and J.~Fisher, ``Logic programming to
  predict cell fate patterns and retrodict genotypes in organogenesis,''
  \emph{Journal of The Royal Society Interface}, vol.~11, no.~98, 2014.

\end{thebibliography}

\end{document}